\documentclass[journal=jacsat,manuscript=article]{achemso}

\usepackage[version=3]{mhchem} 
\usepackage[T1]{fontenc}       

\author{Dalibor \v{S}tys}
\affiliation[ICPF]{University of South Bohemia in \v{C}esk\'{e} Bud\v{e}jovice, Faculty of Fisheries and Protection of Waters, South Bohemian Research Center of Aquaculture and Biodiversity of Hydrocenoses, Institute of Complex Systems, Z\'{a}mek 136, 373 33 Nov\'{e} Hrady, Czech Republic}
\email{stys@jcu.cz}
\phone{+420 38777 384}
\author{Petr Jizba}
\affiliation[FJFI]{Faculty of Nuclear Sciences and Physical Engineering, Czech Technical University, B\v{r}ehov\'{a} 7, Prague~1, Czech Republic}
\author{Anna Zhyrova}
\affiliation[ICPF]{University of South Bohemia in \v{C}esk\'{e} Bud\v{e}jovice, Faculty of Fisheries and Protection of Waters, South Bohemian Research Center of Aquaculture and Biodiversity of Hydrocenoses, Institute of Complex Systems, Z\'{a}mek 136, 373 33 Nov\'{e} Hrady, Czech Republic}
\author{Renata Rycht\'{a}rikov\'{a}}
\affiliation[ICPF]{University of South Bohemia in \v{C}esk\'{e} Bud\v{e}jovice, Faculty of Fisheries and Protection of Waters, South Bohemian Research Center of Aquaculture and Biodiversity of Hydrocenoses, Institute of Complex Systems, Z\'{a}mek 136, 373 33 Nov\'{e} Hrady, Czech Republic}
\author{Kry\v{s}tof M. \v{S}tys}
\affiliation[ICPF]{University of South Bohemia in \v{C}esk\'{e} Bud\v{e}jovice, Faculty of Fisheries and Protection of Waters, South Bohemian Research Center of Aquaculture and Biodiversity of Hydrocenoses, Institute of Complex Systems, Z\'{a}mek 136, 373 33 Nov\'{e} Hrady, Czech Republic}
\author{Tom\'{a}\v{s} N\'{a}hl\'{i}k}
\affiliation[ICPF]{University of South Bohemia in \v{C}esk\'{e} Bud\v{e}jovice, Faculty of Fisheries and Protection of Waters, South Bohemian Research Center of Aquaculture and Biodiversity of Hydrocenoses, Institute of Complex Systems, Z\'{a}mek 136, 373 33 Nov\'{e} Hrady, Czech Republic}

\title[Hotchpotch model of the Belousov--Zhabotinsky reaction]
  {Multi-state stochastic hotchpotch model gives rise to the observed mesoscopic behaviour in the non-stirred Belousov--Zhabotinsky reaction}


\begin{document}







\begin{abstract}
Mesoscopic dynamics of self-organized structures is the most important aspect in the description of complex living systems. The Belousov--Zhabotinsky (B--Z) reaction is in this respect a convenient testbed because it represents a prototype of chemical self-organization with a rich variety of emergent wave-spiral patterns. Using a multi-state stochastic hotchpotch model, we show here that the mesoscopic behaviour of the non-stirred B--Z reaction is both qualitatively and quantitatively susceptible to the description in terms of stochastic multilevel cellular automata. This further implies that the mesoscopic dynamics of the non-stirred B--Z reaction results from a delicate interplay between a) a maximal number of available states within the elementary time lag (i.e. a minimal time interval needed for demise of a final state) and b) an imprecision or uncertainty in the definition of state. If either the number of time lags is largely different from 7 or the maximal number of available states is smaller than 20, the physicochemical conditions are inappropriate for a formation of the wave-spiral patterns. Furthermore, a white noise seems to be key for the formation of circular structures (target patterns) which could not be as yet systematically explained in existing models.
\end{abstract}

\section{Introduction}
During the last two decades, a number of new paradigms for understanding complex living systems have emerged. These include, e.g. theory of dynamical systems, theory of complexity, nonlinear dynamics, evolutionary physics, and critical phenomena\cite{Pikovsky2001, Tiezzi2006}. Among these, chaotic attractors, (multi-)fractals, self-assemblies, dissipative structures and self-organization represent some of the most promising recent concepts. A particularly important testbed for a conceptualization of pattern formation in self-organizing systems is the B--Z reaction\cite{Belousov1959, Zhabotinsky1964, Biosa, Belmont, Kitahara, Liveri, Rustici} which belongs among the most extensively studied examples of chemical self-organization. However, despite decades of intensive research, there are still ongoing controversies over the actual chemical kinetics (i.e. details of rates of chemical reactions involved) and the mesoscopic dynamics (i.e. exact nature and mechanism of patterns formation at the mesoscopic scale) of the B--Z reaction\cite{cross, Burdoni, Holley}.

The B--Z reaction is considered as a textbook example of the so-called excitable medium\cite{Neumann, Kinoshitaetal2013, Dewdney1988, Wilensky2002}. Majority of available chemical models which aim to explain the chemical self-organization are based on the standard reaction-diffusion analysis and the law of mass action applied to a few selected reactions\cite{cit1,Ortiz2006}. On the other hand, self-organization is a hallmark of a far-from-equilibrium dynamics which appears difficult to reconcile with a common-sense chemical reaction scheme based on the law of mass action.

Turing patterns that appear in some reaction-diffusion models are often considered as a theoretical embodiment of the B--Z patterns\cite{Turing1955}. This is wrong for at least three reasons: Turing patterns (a) can explain only wave B--Z patterns but not spirals\cite{cross}, (b) appear only at specific parameter values in the reaction-diffusion equations, therefore they are unstable under fluctuations of parameters (in contrast to experiments where patterns are observably robust), (c) are stable solutions of Turing's reaction-diffusion equations while, in the experiment, we observe a dynamic system on a trajectory through the state space towards a limit cycle with alternating spirals and wave fragments.

The mesoscopic description of the B--Z reaction is typically modeled with a cyclic cellular automaton which often generates patterns similar to the B--Z wave patterns found near to the final stage of the reaction. The morphological characterization of patterns is pivotal in these approaches while chemical aspects are often of a secondary interest.

So far, mesoscopic studies of the B--Z reaction have been limited to a low-level cellular automata\cite{Hiltemann2008} which can reasonably well account only for some of the observed B--Z wave patterns, while it is as yet unclear how to generalize these approaches to obtain a full-fledged evolution of wave patterns together with dynamics of spiral patterns. To the best of our knowledge, the influence of the number of levels has been systematically studied only in one case\cite{Hiltemann2008} while most of the systematic studies in the literature have been confined to maximally 8 levels\cite{Wuensche2011}.

In order to numerically implement a hotchpotch model, we have adopted an approach based on the version of Wilensky NetLogo model\cite{Wilensky2002}. In our case the model is limited to 200 achievable state levels and simulated on a square 50 $\times$ 50 grid. After a random setup of the space distribution of initial centers of $state(t = 0) \in [0, maxstate]$ as
\begin{equation}
state(t=0) = \mbox{random}(maxstate+1),
\label{Eq1}
\end{equation}
where $maxstate$ is the maximally achievable number of levels of the cell state. The model at each time step $t+1$ may proceed in four possible ways:
\begin{enumerate}
\item When a cell is at the $state(t) = 0$, so-called {\em quiescent}, it may be ``infected'' by surrounding cells according to the equation
\begin{equation}
state(t+1) = \mbox{int}\left(\frac{a}{k_{1}}\right) +\mbox{int}\left(\frac{b}{k_{2}}\right),
\label{Eq2}
\end{equation}
where $a$ and $b$ is a number of cells at the $state \in (0, maxstate)$ and $state = maxstate$, respectively, $k_{1}$ and $k_{2}$ are characteristic constants of the process and $maxstate$ is a maximum allowed level of state/excitation.
\item When a cell is at the $state(t) \in (0, maxstate)$, its new state is calculated as
\begin{equation}
state(t+1) = \mbox{int}\left(\frac{\sum_{n=1}^{8} state_{n} (t)}{a + b +1} + g\right),
\label{Eq3}
\end{equation}
where $state_{n} (t)$ is a state of the $n$-th cell in the Moore neighbourhood, which directly surrounds the examined cell, and $g$ is another arbitrary constant. For simplicity's sake, we denote in the following text  $\frac{\sum_{n=1}^{8} state_{n} (t)}{a + b +1}$ and $g$ as 2a and 2b, respectively.
\item When a cell is at the $state(t) > maxstate$, then
\begin{equation}
state(t+1) = maxstate.
\label{Eq4}
\end{equation}
\item When a cell achieves the $state(t) = maxstate$, then
\begin{equation}
state(t+1) = 0.
\label{Eq5}
\end{equation}
\end{enumerate}

Here we posit that our improved stochastic version of the multilevel hotchpotch model not only faithfully represents the dynamics of wave-spiral patterns of the B--Z reaction but also provides insight into underlying chemical reactions in terms of the Eyring activated complex theory\cite{Eyring1935}. The emergence of correct spatial structures in our model depends on the ratio between the number of available states within the elementary time lag and on the rate of the internal increase in excitation process. At appropriate levels of noise, we observe a dynamical evolution of the system through circular waves up to the final stage --- spiral-wave interchange. This structure-stabilizing mechanism via the noise is a typical trademark of a mesoscopic description and is not simply attainable via microscopic deterministic rules. By the presence of the noise, we can explain many controversies in self-organization in the Nature.

\section{Results}

\subsection{Modelling the Belousov--Zhabotinsky reaction in excitable media and constructive role of noise}

The B--Z reaction is not easily comprehensible in terms of the standard law of mass action (which represents the ``canonical method'' of interpretation of the chemical reactivity). This is due to the fact that the reaction space is separated into regularly evolving/travelling structures and that one has to consider a large number of interlocked chemical processes involved. In this work, we report on a new cellular-automaton based stochastic model of the B--Z reaction. The model retains some of the key features of the multi-level hotchpotch machine which, however, outperforms both in its ability to faithfully mimic the onset stage of the B--Z reaction and its potential to correctly describe the morphology of the interacting wave-spiral patters during their evolution.

Figure 1a compares a late stage of the B--Z reaction (full data are accessible in Supplementary Material 1) in our least possible spatial constraining (a 200-mm Petri dish) and roily  conditions (gentle mixing at 1400 rpm using an orbital mixer) with simulation of the modified Wilensky model of excitable medium\cite{Wilensky2002}. The structures are, for some parameter ranges, astonishingly similar. However, the most regular spirals and waves, best comparable to the model, are expected to arise in a very gently mixed, homogenous solution of thin layer in a vessel of the unlimited size which does not spatially constrain the evolution of waves.

In order to achieve the manifest morphological similarity between the B--Z experiments and our simulation, we implemented the following changes into the Wilensky model:
\begin{enumerate}
\item the enlargement of the cellular grid to 1000 $\times$ 1000,
\item start from a very few points which enabled to analyze the behaviour of individual centers of emanation,
\item a sequence of switching the values of cell states from natural to decimal numbers which extended the span of each cellular state,
\item the addition of a uniform white noise to each automaton step which compensated for our limited knowledge of precise underlying mechanism, and
\item the extension of the number of achievable states $maxstate$ and rate of the internal cell excitation $g$ up to 2000 and 280, respectively, to smooth the model waves.
\end{enumerate}

The first modification --- usage of the finer grid --- suppressed the influence of the nonidealities of the periodic boundaries on the evolution of the model system. The second intervention into the Wilensky model was performed using random-exponential function for generation of the starting (ignition) points. Multiplication of each cell state by the $meanPosition$ of the exponential distribution, i.e.
\begin{equation}
state(t=0) = \mbox{random-exponential}[(maxstate+1) \times meanPosition]\, ,
\end{equation}
ensured that the simulation started with a small number of the ignition points.

At this initial condition, we first tested the influence of the $states(t+1)$ of natural numbers due to the rounding in processes 1--2 (see Eqs.~(\ref{Eq2})-(\ref{Eq3})), thus new states in time $t+1$ were calculated as
\begin{equation}
state(t+1) = \mbox{round}\left(\frac{a}{k_{1}}\right) +\mbox{round}\left(\frac{b}{k_{2}}\right),
\end{equation}
and
\begin{equation}
state(t+1) = \mbox{round}\left(\frac{\sum_{n=1}^{8} state_{n} (t)}{a + b +1} + g\right).
\end{equation}
This modification, which was originally implemented to start the process from these few centers (ignition points) quite surprisingly increased the structural-patterns similarity between the B--Z experiment and the simulation. The results are shown in~Figure 1b--c and~Supplementary Materials 2--4 and were discussed in some detail in\cite{Stysetal2015a}. In~Figure 1b, we present early simulation steps 2, 4, 14, and 16 after the ignition in process 1. For $k_{1} = 3$ and $k_{2} = 3$, at least two non-zero points in a proper configuration $a$, $b$ were required for the evolution of the waves in the simulation, since at least one addend in process 1 has to be equal to 1. In this case, the early evolution gave octagons (Figure 1b, i) and final state was formed by spirals (Figure 1c, i). In contrast, if $k_{1} = 2$ and $k_{2} = 2$, then, e.g. $state (t+1) = \mbox{round}(\frac{1}{2}) + \mbox{round}(\frac{0}{2}) = 1$ and the non-zero cell was surrounded by evolving wave of 8 cells in $state (t+1) = 1$. This early evolution resulted in squares with central circular objects (Figure 1b, ii) which further led to the filamentous structures (Figure 1c, ii).

The next step softened the definition of the state by allowing 1 decimal place in Eqs.~(\ref{Eq2})-(\ref{Eq3}), i.e.
\begin{equation}
state(t+1) = \mbox{precision}\left(\frac{a}{k_{1}}\right) +\mbox{precision}\left(\frac{b}{k_{2}}\right),
\end{equation}
and
\begin{equation}
state(t+1) = \mbox{precision}\left(\frac{\sum_{n=1}^{8} state_{n} (t)}{a + b +1} + g\right),
\end{equation}
and kept $k_{1} = 3$ and $k_{2} = 3$ (Supplementary Material 5). If $state(t+1) = \frac{1}{3} + \frac{0}{3} = \frac{1}{3}$, the first layer of 8 cells in $state (t+1) = \frac{1}{3}$ evolves in the Moore neighbourhood. This modification provided the trajectory as observed for $k_{1} = 2$ and $k_{2} = 2$ with natural number states. More decimal places did not change the state space trajectory any more. This proves that the trajectory is determined by the ratio of the $g$ constant to the number of levels $maxstate$. The $g/maxstate$ ratio close to 7 (28/200) showed the highest structural similarity to the experiment. There seems to be also a lower limit for the proper development of the system trajectory. In case of $g = 1$ and $maxstate = 7$, we observed evolution of square spirals and the simulation grid was never fully covered by spirals and waves. At the 2/14 ratio, the grid was already completely covered. At the 3/20 ratio, the circular structures similar to those observed in the experiment were firstly observed. At keeping the $g/maxstate$ ratio, the next increase of the $maxstate$ value smoothed only the edges of the structures. Obviously, 7 state levels does not allow to cover the Moore space with 8 neighbouring cells and 14 state levels are not yet sufficient enough to create an appropriate curvature.

The course of the simulation and the type of the limit set (late state) depends on the type of the {\em Garden of Eden} as follows: a) the emergence of the ignition point needs one or two neighbouring non-zero points and determines the overall type of the trajectory and b) the number and distribution of ignition centers determines the duration of each trajectory phase. There are as many Gardens of Eden as possible geometrical set-ups of the ignition points, however, the thickness of square and circular waves, shapes of the central objects and the structure of the limit set remain the same. The multilevel cellular automata exhibit less versatile trajectory than the low-level ones\cite{Wuensche2011}.

In order to compensate our lack of knowledge of the details of the process, we introduced noise into each sub-process in Eqs.~(\ref{Eq2})-(\ref{Eq3}) by multiplication of the terms $\frac{a}{k_{1}} + \frac{b}{k_{2}}$, $\frac{\sum_{n=1}^{8} state_{n} (t)}{a + b +1}$, and $g$, respectively, by the factor $1 + noise$. The noise was generated by a computer random number generator and its level was upper limited to be the uniform noise  within certain range. At $k_{1} = 3$, $k_{2} = 3$ and precision of 10 decimal places, levels of noise were 6\%, 12\%, and 30\% for $\frac{a}{k_{1}} + \frac{b}{k_{2}}$, $\frac{\sum_{n=1}^{8} state_{n} (t)}{a + b +1}$, and $g$, respectively. When ignition led to circular structures with filled interior, the noise restored spiral formation (Figure 1b--c, iii and Supplementary Material 5) and the simulation was the most similar to the experiment (Figure 2 and Supplementary Material 1). In other experiments at $k_{1} = 2$ and $k_{2} = 2$ (data not shown), we observed that the respective noise levels gave the same trajectory. A formation of circular structures which arose as watersheding of regular structures was also the same as observed for the natural-number states.

The proper combination of $g/maxstate$ values and uniform noise range (Figure 2a) generates a sequence of simulated structures as follows:
\begin{itemize}
\item The simulation grid is filled with systems of square dense waves. This has not been observed in the experiment and we interpret it as a
{\em lag phase}, which precedes the observed formation of circular waves.
\item Circular structures emanate from the centre of square waves.
\item At the certain state, the simulation grid is nearly covered by large circular structures. A few spirals occur and break into a new generation of spirals.
\item Final state is similar to that in the simulation where the states are natural numbers, $k_{1} = 3$ and $k_{2} = 3$ (Supplementary Material 4), however, the waves are about 2 grid elements thicker.
\item In the terminology of the multilevel cellular automata\cite{Wuensche2011}, the uniform (white) noise shifts the system to the basin of attraction similar to the system where only natural numbers are allowed and open a novel system trajectory through the state space.
\end{itemize}

Let us mention further key similarities between our simulation and actual experiments (Figure 2b--c):
\begin{itemize}
\item The chemical waves do not interfere like material waves but merge.
\item The chemical waves do not maintain the shape (as is the case, e.g., for solitons\cite{Blasone:2011}).
\item The morphology of interacting patterns (merger patterns) in simulations has comparable traits as in real experiments.
\item Quantitative features of the limit sets, i.e., the last evolutionary stage of the wave-spiral patterns can be set
as close as possible to actual experimental data by an appropriate choice of the parameter range.
\end{itemize}

It should be stressed that the possibility to replace partly the elementary anisotropy by the uniform noise (see Figure 4) --- the constructive role of the noise --- is a novel observation, which can be viewed as a superposition of starting conditions.

\subsection{Decay and re-assembly of transition state complex and theory of mesoscopicity in chemical systems}

When noise matters, an observed process is typically mesoscopic. It does follow neither the deterministic rules of the microscopic (or purely mechanical) system nor the statistical-physics tenet of Boltzmann that only the most frequent events are observed\cite{cit1}.
So, in case of the B--Z reaction, a chemical origin of mesoscopicity should be sought.

The key process at the end of the reaction scheme is the breakage of a crucial chemical bond in the Eyring transition state complex, possibly the decomposition of brominated malonic acid to formic acid and carbon dioxide\cite{Fieldsetal1973, RoviskyZhabotinsky1988}. A sketch of molecules involved in the last reaction step is outlined in~Figure 3. This process is hidden in the experiment in which we observe only the reduction of the Fe$^{3+}$-phenanthroline complex. The rest of, known and unknown, reactions in all their reaction intermediates contributes to the restoration of this activated complex. The activated complex consequently reduces Fe$^{3+}$ which enables our measurement. The scheme described here still includes breakage of a few individual bonds, possibly in certain orders.  Moreover, it is not certain that this redox process is the bottleneck process. It should be also noted that there is a deep controversy about the actual details of the mechanism including the role of molecules involved in the carbon dioxide evolution. Thus the proposed chemical mechanism must be understood more as illustration of the complexity and of the origin of time extent of key reaction steps. The chemical reaction is certainly not occurring at timeless instant as it is assumed in the reaction--diffusion model.

We propose that the state changes of the activated complex are described by processes 2--3 (see Eqs.~(\ref{Eq3})-(\ref{Eq4})). Process 4 (see Eq.~(\ref{Eq5})) represents the breakage of the complex, which vacates the space for the restoration of a new activated complex in process 1 (see Eq.~(\ref{Eq2})). The mesoscopicity may be explained from the multitude of processes involved in these elementary steps.

To examine properties of the elementary spatial unit (pixel in the simulation), we analyzed the properties of process 2. We assume that process 2 (see Eq.~(\ref{Eq3})) reflects all reorganization processes and reactions of all parts of the complex, down to the level of a bond which breaks upon the disruption of the complex. We split the process into sub-processes 2a and 2b. Process 2a is the sum of all processes which occur between the elementary spatial unit and the immediate surroundings. Larger surroundings were ignored, since their influence was assumed to be mediated by neighbouring elementary units. Process 2b represents all processes within this unit. The relatively high noise level of 0.14 and 0.30 for process 2a and 2b, respectively, includes our ignorance of the details and randomness of the process.

Figure 4a and Supplementary Material 6 show the results of the noise-free simulation at $maxstate = 2000$, $k_{1} = 3$, $k_{2} = 3$ and different $g$. At the $g/maxstate = 280/2000$, the process exhibited the same trajectory as at the $g/maxstate = 28/200$. At the $g/maxstate = 10/2000$, the process was in its initial part similar to the course at the $g/maxstate = 28/200$, $k_{1} = 2$ and $k_{2} = 2$, later, it tended to give round structures. At the $g/maxstate = 1000/2000$, the course of the simulation was very different from any previous simulation. The final state of this simulation was characterized by spiral doublets --- ram's horns\cite{Fischetal1991}. At the $g/maxstate = 1/2000$, the process was already very diffusive. Obviously, the ratio between the constants of processes 2a and 2b determines the geometry of the process. For the course of the simulation similar to the experiment, the suitable $g/maxstate$ ratio is equal to 7. The values of constants $g$ and $maxstate$ themselves do not significantly influence the course of the simulation.

The time needed for the decay of the active state complex and its restoration in process 4 determines the characteristic timescale of the whole simulation. The ratio $g/maxstate$ determines properties of waves and the size of the spatial unit. This ratio represents an aliquot of energy/entropy needed for the decay of the transition state complex in process 2b. In order to observe waves, process 2b needs to be significantly faster than the transition from the neighbourhood in process 2a and, also, the neighbourhood has to be asymmetric. In other words, noise represents geometric asymmetry (2a noise) and internal kinetic irregularity (2b noise) of the neighbourhood.

\subsection{Size of the elementary spatial unit}

Figure 5 shows the analysis of wave profiles in the B--Z experiment.  The striking similarity of the simulation to real experiment intensity profiles of dense waves (Figures 1, 2 and 5) motivated us to guess the number of molecules per an elementary spatial unit (i.e. the pixel of experimental wave). The number of elementary units per the width of the wave was in the range of 10--20. Since the average width of the wave was 1.5 mm, the elementary unit had 0.07--0.15 mm. The solution above the elementary unit had thickness and volume of 0.5 mm and 10$^{-2}$ mm$^{3}$, respectively. Then, the solution contained ca. 3 $\times$ 10$^{13}$ and 10$^{10}$ molecules of water and reactants per elementary unit, respectively. This number lies within the thermodynamic limit. The source of the mesoscopicity has to be sought in the physico--chemical dynamics. It means that only a few energetic/re-organizational events occur within a given time. Since an elementary spatial unit contains roughly 10$^{10}$ molecules of reactants, it is probably unsatisfactory to explain the energetic transition only due to one Eyring-type reaction complex. It is likely that we are dealing with a phase separation which gives rise to structures of an analogous type as, e.g., in liquid crystals\cite{Blasone:2011}.

\section{Discussion}

Both wavy and spiral patterns often occur in the Nature. However, despite being usually formed in continuous media, they can also be generated artificially in the threshold-range of cyclic multilevel cellular automata\cite{Wuensche2011}. The generation of waves and spirals in
cellular automata has been initially perceived by the scientific community with scepticism\cite{Lauzeraletal1997}. The reasons why the formation of spirals by cellular automata is not easily accepted by general chemical community is a) the enforcement of the natural number states (i.e., a coarsening of the configuration space) b) and the need of discrete dynamics on a regular grid which is driven by deterministic albeit not differential rules. In our simulation, the former limitation is completely suppressed by a (near--)continuous set of levels. The latter limitation seems to be only a prejudice. Whereas the formation of B\'{e}nard cells in Rayleigh--B\'{e}nard convection  is driven by temperature gradient\cite{Benard} and the speed gradient generate cells in the Taylor--Couette flow, the Gibbs energy gradient drives the discrete cells formation in the B--Z reaction.

The basic principles of the formation of waves and spirals, i.e. the specific mesoscopic behaviour, are in our simulation following:
\begin{itemize}
\item Proper $g/maxstate$ ratio for a given spatial discretization\cite{Stysetal2015a}. A high and low $g/maxstate$ ratio promotes structural asymmetry and diffussiveness, respectively.
\item Presence of sufficient amount of irregularities due to other processes of comparable rates and leads to a noise, which is specific to each (sub)-process. Only at the appropriate combination of noise in processes 2a and 2b, we observed concentric circular waves followed by spirals. When the specific noise, given by decimal places in state levels, by the pulsed white noise, we may conclude that an increase of the noise in process 2a -- i.e. increase of the diffusiveness -- leads to the faster formation of spirals without any circular waves. The asymmetry of the surroundings of the elementary unit, as observed in the simulation with natural number state levels, $k_{1} = 3$ and $k_{2} = 3$, is probably a sub-set of the appropriate specific noises modulating the wavefront. We call the whole range of state-level distributions, which attenuate the specific mesoscopic behaviour, a mesoscopic noise.
\end{itemize}

Main differences between the experiment and the simulation are following:
\begin{enumerate}
\item In the simulation, formation of a few dense waves precedes the emanation of circular waves (never observed in the experiment).
\item In the experiment, four different wave frequencies are observed (not yet achieved in the simulation).
\item In the simulation, dense waves/spirals are arising from remnants of broken initial dense waves. In the experiment, we observe a formation of new centers of dense waves, often at place of evolving micro-bubbles. But scenario of broken waves may also be found by careful inspection of the experiment.
\item In the experiment, after the state of dense waves/spirals, the system still evolves. The waves broaden due to the exhaustion of chemical energy (see a re-started experiment in Supplementary Material 1).
\end{enumerate}

Many differences may be attributed to our ignorance of the proper character of the underlying mesoscopic noise. The introduction of two noise levels into the simulation indicates that the mesoscopic noise originates from two different physico-chemical processes which lead to the geometrical and kinetic asymmetry.

Even if we use the square tessellation and neighbourhood which allows to cover the whole simulation grid --- the Moore neighbourhood, the simulation exhibits the formation of {\em circles}. A symmetrical ignition rule gives rise to circles and filamentous structures, whereas the simplest asymmetrical ignition leads to interchanges of spirals and waves (with fractal-like structure). The noise provides a proper condition for switching between these two tendencies. The question remains whether the same rules of ignition and noise introduction give rise to the same geometrical structures in any tessellation or whether the neighbourhood of the active complex is, for some reason, due to the physico-chemical, or a more fundamental physical or geometric, principle, always a Moore-like neighbourhood. We can imagine some organization of molecules into a near-regular square lattice which is maintained by chemical dynamics (i.e., by movement of the molecules), similar to that by which the B\'{e}nard cell arises in thin layers of viscous liquids.

We assume that the time is discretized by a step needed for the bottleneck process, i.e. for the restoration of the original concentrations of molecules in spatial element after their depletion. It is described by processes 3--4 (see Eqs.~(\ref{Eq4})-(\ref{Eq5})) and represented by 2 time elements. All other processes are related to this time measure. Thus, for the formation of the proper structures, a near regular structure (lattice) where the processes of depletion and repletion take 7 and 2 time elements, respectively, is needed. In the noise-free hotchpotch machine, mixtures of spirals and waves are formed in a relatively broad range of the $g/maxstate$ ratio\cite{Stysetal2015a} -- between $200/2000$ and $800/2000$. All these arguments, including the trajectory of any described multilevel cellular automaton, explain the great similarity to the reaction.

Let us conclude that a computer-based reaction-diffusion model of the B--Z reaction is calculated on a regular grid and all but the fastest process are represented by a stepwise increase/decrease. From this point of view, any reaction-diffusion model is a special kind of the cellular automaton\cite{Fischetal1991, Kapral1991, Wuensche2011}. Apart from a few special cases when an analytic solution of Turing patterns may be used\cite{Turing1955}, the threshold-range cellular automaton offers promising and very realistic tool for description and modelling of the chemical self-organization.

\begin{figure}
\includegraphics[width=\textwidth]{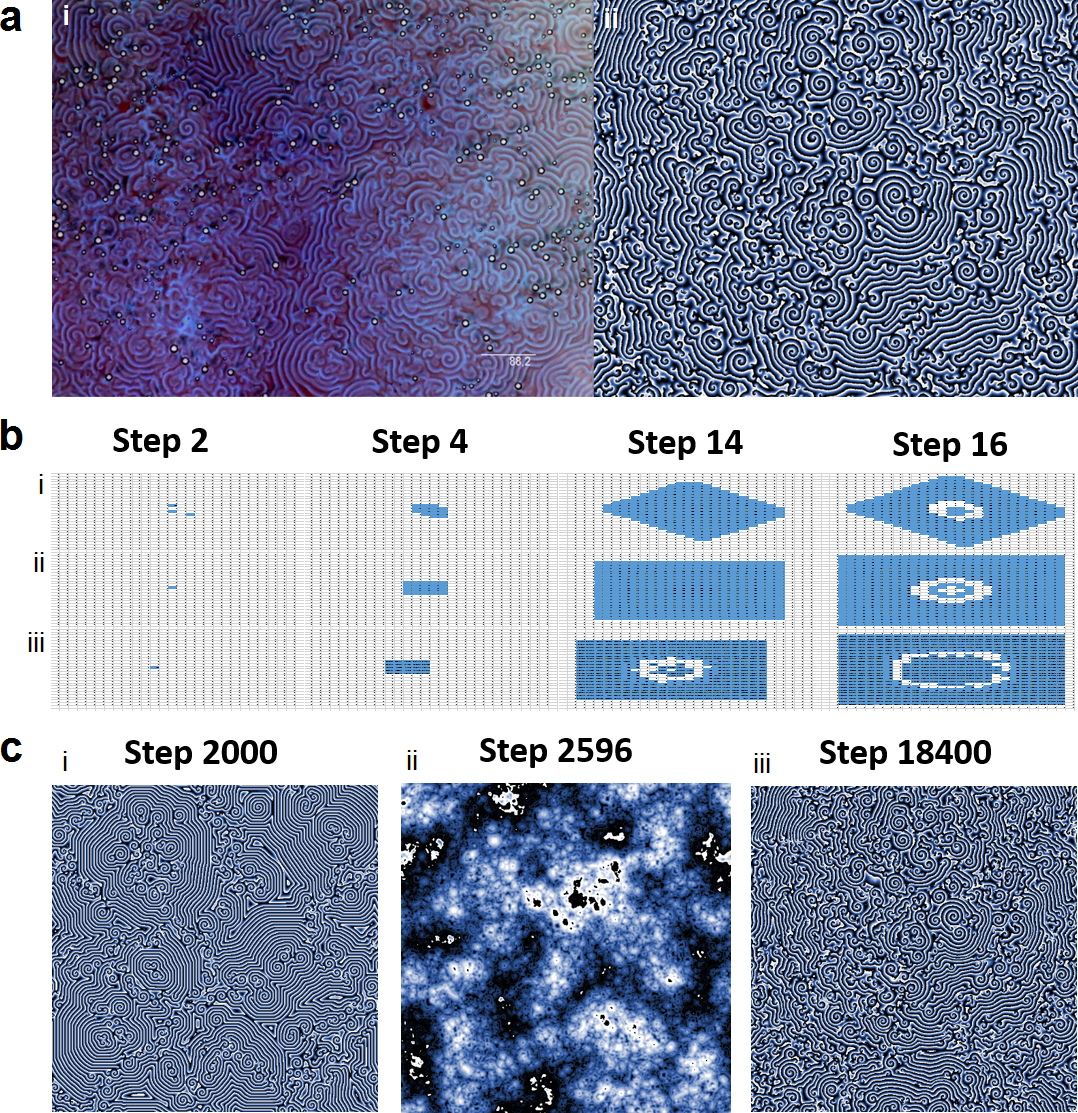}
\caption{Illustration of the key aspects of the B--Z model. (\textbf{a}) Comparison of the B--Z experiment (\textbf{i}) with the simulation at optimal levels of noise for each process, i.e. 9\%, 14\% and 30\% for process 1, 2a, and 2b, respectively, and $k_{1} = 3$ and $k_{2} = 3$ (\textbf{ii}). Images were expanded so as to have comparable widths of traveling waves. (\textbf{b}) Starting points of the simulations (steps 2, 4, 14, 16). The noise-free simulation with natural number states, $k_{1} = 3$ and $k_{2} = 3$ in step 2,000 (\textbf{i}), the noise-free simulation with natural number states, $k_{1} = 2$ and $k_{2} = 2$ in step 2,596 (\textbf{ii}) and the process described under \textbf{a} in step 18,400 (\textbf{iii}). (\textbf{c}) Final states (limit sets) of processes defined in \textbf{b}. For all processes, $g = 28$ and $maxstate = 200$. In the simulation, the black and white corresponds to 0 and $maxstate$, respectively. Original datasets supplied in Supplementary Material 3.
\newline The intensity is colour-coded in the blue colour, the darkest blue represents value 200, the white colour the intensity 0. The unquestionably inspection of the data has to be done using the original data matrices as demonstrated in Fig. 4. }
\end{figure}

\begin{figure}
\includegraphics[width=\textwidth]{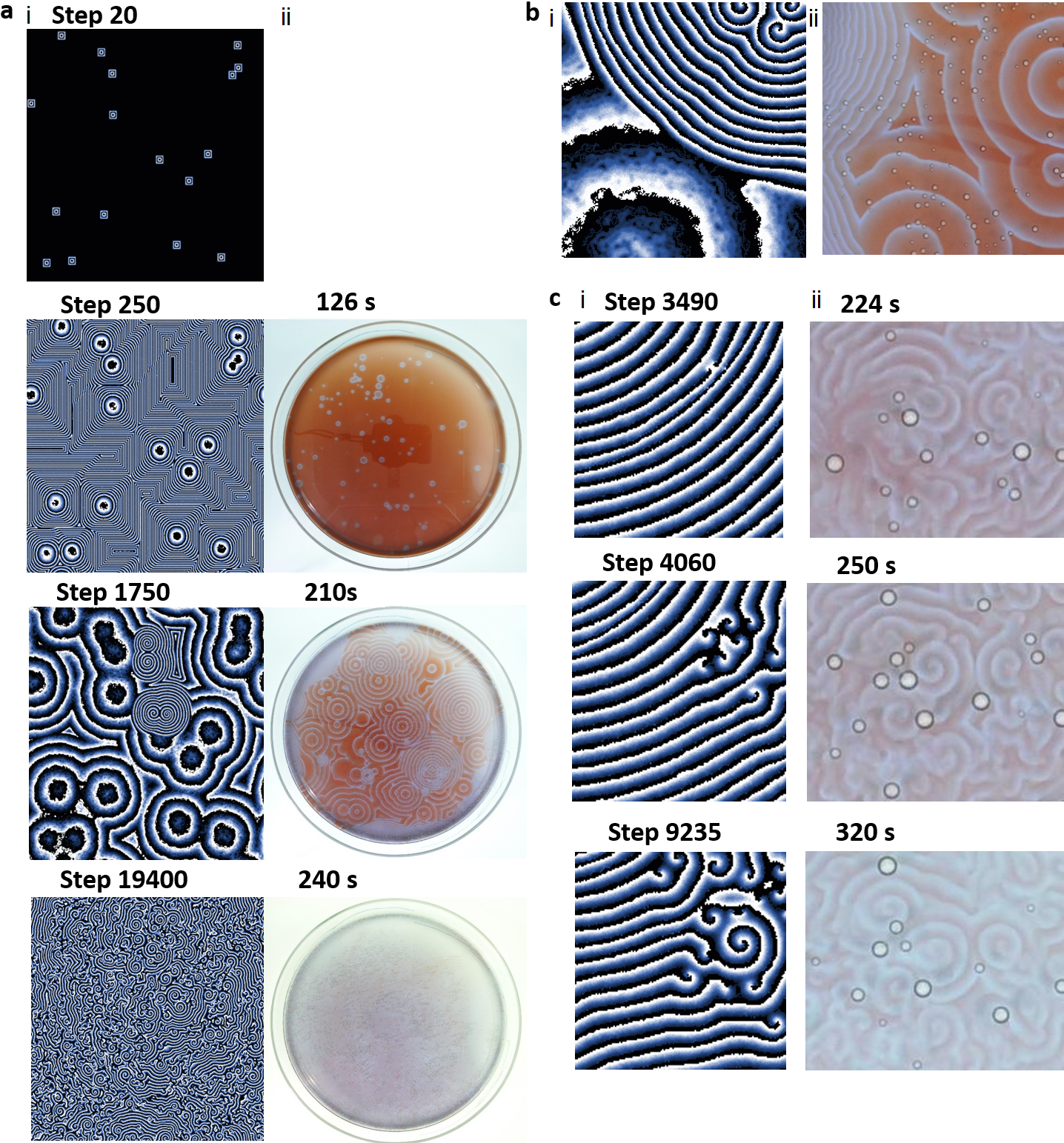}
\caption{Similarities between the trajectories of the simulation at optimal noise and those of the Belousov--Zhabotinsky experiment. (\textbf{a}) Selected states of the simulation (\textbf{i}) and corresponding images from the course of the experiment (\textbf{ii}). The early stage of the experiment corresponds to the lag phase of the experiment when no waves evolve. For the later stages of the simulation, corresponding structures were found in the experiment. (\textbf{b}) Sections of images which show wave merging. Similar behaviour was not found for material waves and another wavelike structures and indicates that threshold-range cellular automata (\textbf{i}) are proper models for observed phenomena in the Belousov--Zhabotinsky experiment (\textbf{ii}). (\textbf{c}) States of spiral formation. In the simulation (\textbf{i}), the distortion of the dense waves leads to their merging which is the source of formation of spirals. In the experiment (\textbf{ii}), the source of the distortion is often a bubble of carbon dioxide. Otherwise, the formation of spirals is similar to the experiment. For all processes, $g = 28$, $maxstate = 200$, $k_{1} = 3$ and $k_{2} = 3$. In the simulation, the black and white corresponds to 0 and $maxstate$, respectively.}
\end{figure}

\begin{figure}
\includegraphics[width=0.6\textwidth]{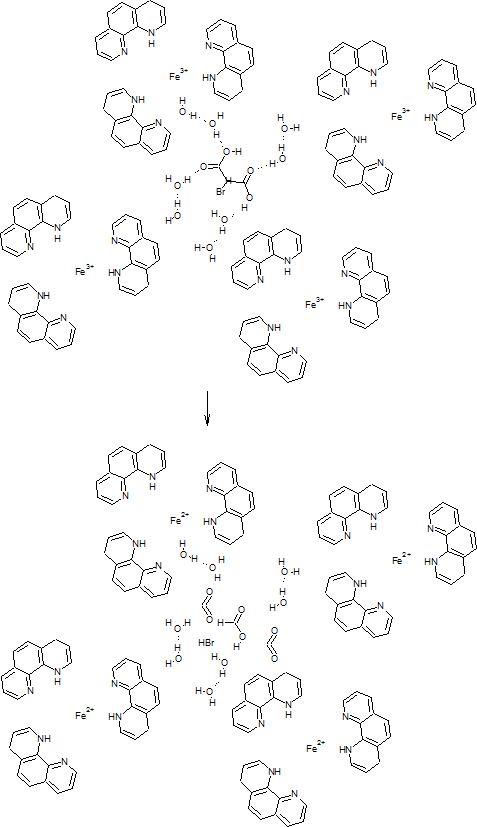}
\caption{Eyring transition state complex in the Belousov--Zhabotinsky reaction. Selection of crucial molecules involved in oxidation of a final bond is shown. During the reaction, a few bonds are broken. Vibration mode of the bond which is broken first is responsible for the conversion of the activated complex to the product\cite{Eyring1935}. In the experiment, the reduction of Fe$^{3+}$-phenanthroline complex to the Fe$^{2+}$-phenanthroline complex is monitored. The absorbance of the Fe$^{2+}$ complex is negligible in comparison to the Fe$^{3+}$ form\cite{Kinoshita2011}.}
\end{figure}

\begin{figure}
\includegraphics[width=\textwidth]{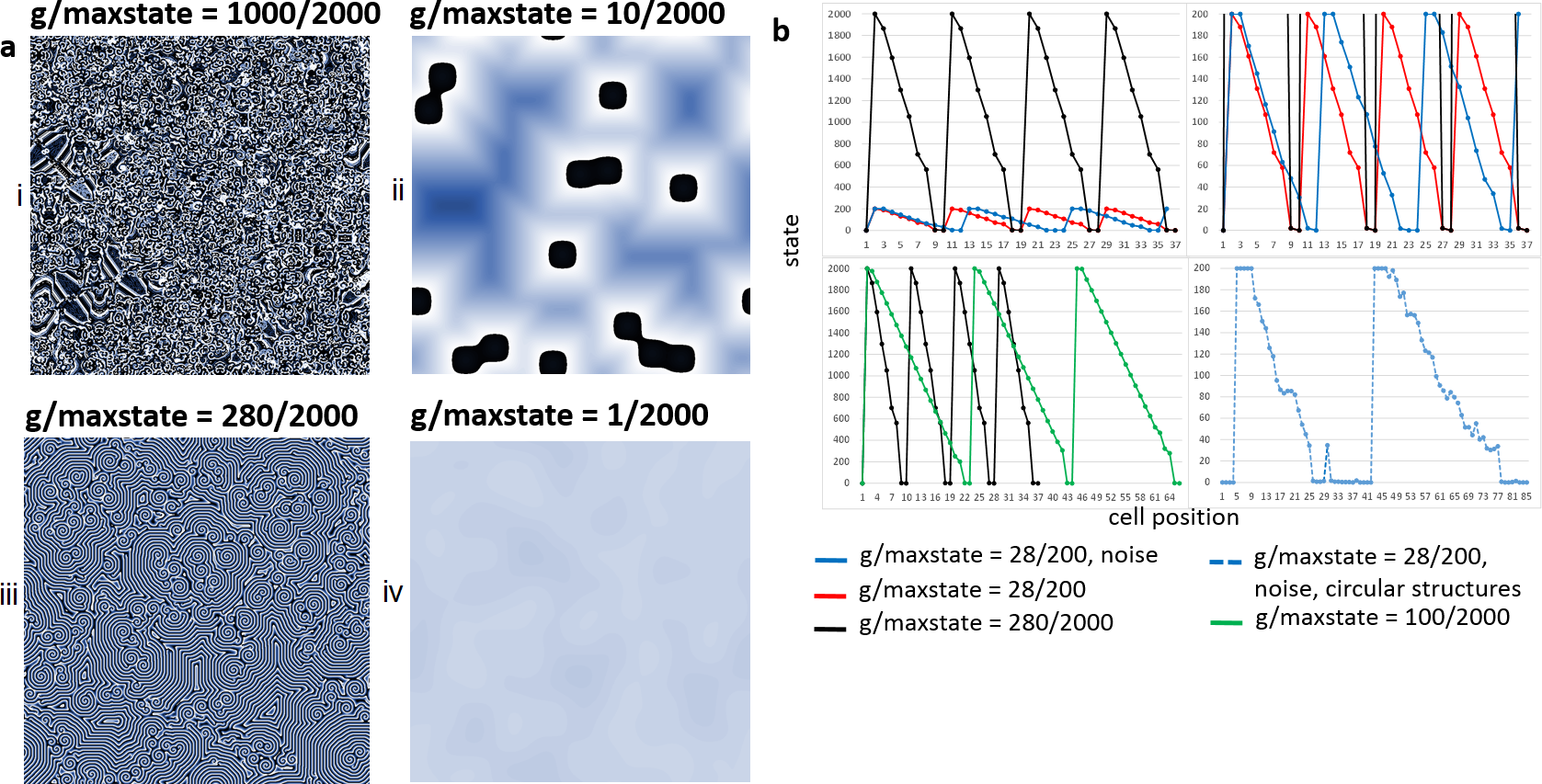}
\caption{Ratio of processes 2a and 2b determines the size of elementary unit and the type of state trajectory. (\textbf{a}) Images of later states of simulation at different $g/maxstate$ ratios, $k_{1} = 3$, $k_{2} = 3$ and $noise = 0$. At $g/maxstate = 1000/2000$ (\textbf{i}), spirals evolve into forms of ram's horns. To the opposite, $g/maxstate = 10/2000$ (\textbf{ii}) does not form spirals. At $g/maxstate = 1/2000$ (\textbf{iv}), the process is fully diffusive. At $g/maxstate = 280/2000$ (\textbf{iii}), the trajectory is almost identical to the experimental trajectory. (\textbf{b}) The intensity profiles of waves at different $g/maxstate$ ratios. Decrease of the $g/maxstate$ ratio leads to the broadening of waves. The intensity profile of the circular structure is very noisy. In the simulation, the black and white corresponds to 0 and $maxstate$, respectively.}
\end{figure}

\begin{figure}
\includegraphics[width=\textwidth]{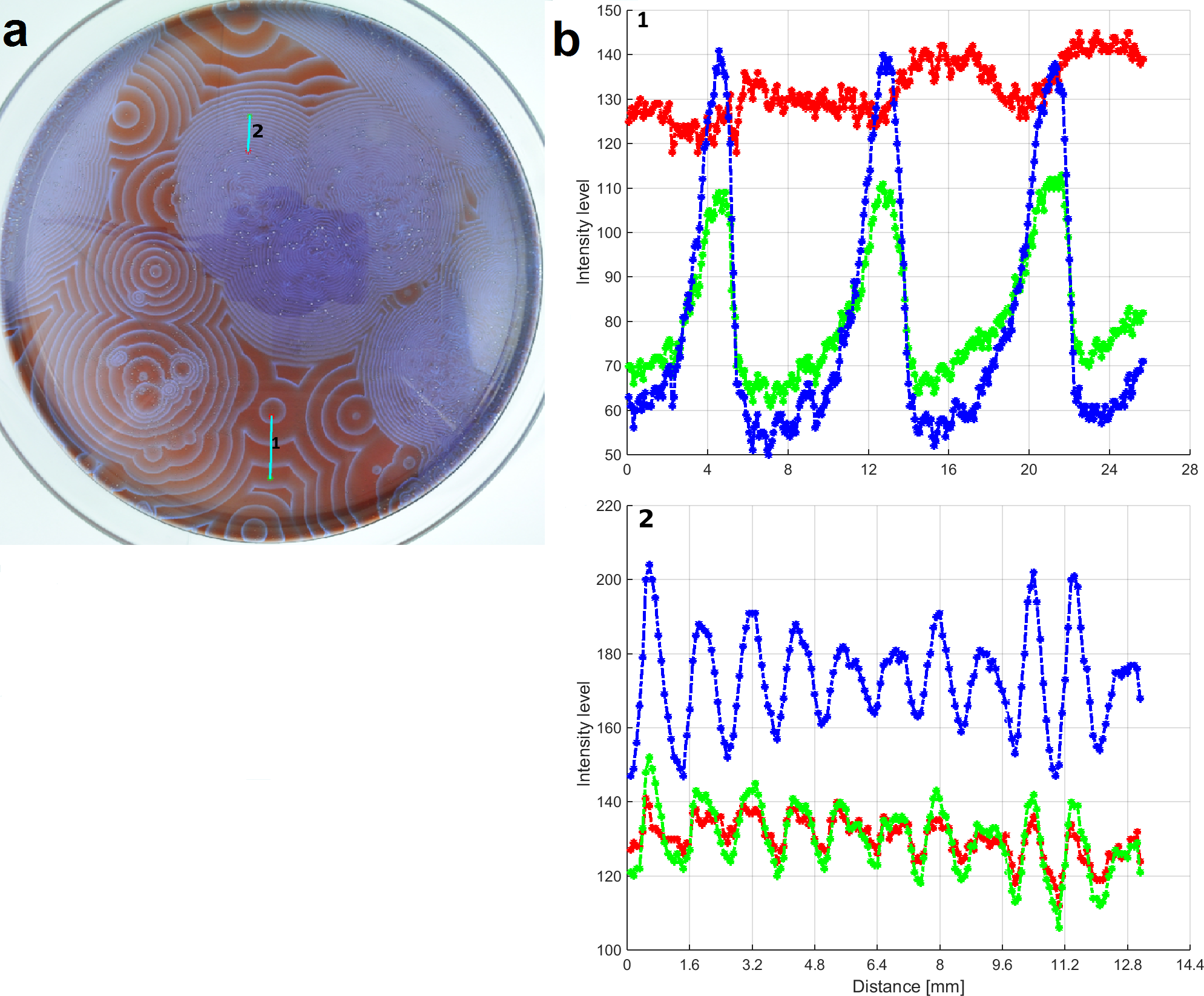}
\caption{Analysis of traveling waves in the Belousov--Zhabotinsky experiment. (\textbf{a}) Figure with identified wave profiles. (\textbf{b}) Intensity profile of the early circular wave (1) and later dense wave (2). Three colours represent camera channels.}
\end{figure}

\section{Experimental}

\subsection{Performance of the chemical reaction}

For experiments, the oscillating  bromate-ferroin-bromomalonic acid reaction (the B--Z reaction recipe\cite{Cohen2010} provided by Dr. Jack Cohen) was chosen. The reaction mixture included 0.34-M sodium bromate, 0.2-M sulphuric acid, 0.057-M sodium bromide (all from Penta), 0.11-M malonic acid (Sigma-Aldrich) as substrates and a redox indicator and 0.12-M 1,10-phenanthroline ferrous complex (Penta) as a catalyst. All reagents were mixed by hand directly in a dish -- a circular Petri dish with diameter of 35, 90, 120 and 200 mm and square dish with side of 75 and 30 mm, respectively ---  in the above-mentioned sequence for 1 min. A special thermostat, which was constructed from a Plexiglas aquarium and a low-temperature circulating water bath-chiller, was keeping a reaction temperature at $26^{\circ}$C.

The chemical waves were recorded by a Nikon D90 camera in the regime of Time lapse (10 s/snapshot) with exposure compensation $+\frac{2}{3}$ EV, ISO 320, aperture $\frac{f}{18}$ and shutter speed $\frac{1}{10}$ s. The original 12-bit NEF raw image format was losslessly transformed to the 12-bit PNG format. See Supplementary Materials. All results and protocols are freely available (Supplementary Material 3).

\subsection{Model of the Belousov--Zhabotinsky reaction}

The model was derived from that provided in the official release\cite{Wilensky2002} of Netlogo 5.1.0 and run there. The full model is provided and described in Supplementary Material 7.

\begin{acknowledgement}

This work was financially supported by the Ministry of Education, Youth and Sports of the Czech Republic -- projects CENAKVA (No. CZ.1.05/2.1.00/01.0024), CENAKVA II (No. LO1205 under the NPU I program) and The CENAKVA Centre Development (No. CZ.1.05/2.1.00/19.0380). P.J. was supported by the GA\v{C}R Grant No. GA14-07983S.

\end{acknowledgement}

\begin{suppinfo}

The following files are available free of charge:
\begin{itemize}
  \item Supplementary Material 1: Video of the Belousov--Zhabotinsky reaction

The movies videoBZ.avi and videoBZreshake.avi were made from sections of the image series BZexperiment and BZ reshake as described below. In the movies, every 5th frame of the series is shown with the frequency of 10 frames per second.

The image series BZexperiment is a dataset from an experiment in a 200-mm Petri dish. The experiment was preceded by a 2-min preparation period, which included mixing of chemicals, shaking on an orbital shaker and lag period under which no travelling waves evolved. The image series BZreshake is a dataset from the re-shaken, i.e. re-started, experiment in a 200-mm Petri dish. Images were captured in 10s intervals.
  \item Supplementary Material 2: Video of the noise-free simulation -- original excitable medium simulation with spirals

The movie  video\_rounding0\_noise0.avi was made from the simulation when only natural numbers were allowed, no noise was included, $g$ = 28, $maxstate$ = 200, $k_{1} = 3$ and $k_{2} = 3$. In the movie, every 30th simulation step is shown with the frequency of 100 frames per second.

   \item Supplementary Material 3: Original data

All original data for experiments and simulations are available upon request to the authors. We apologise for not placing all datasets freely, because some simulations resulted in files of 250 GB size. Complete datasets of a few TB size will be sent upon request on respective hardware media.

   \item Supplementary Material 4: Video of the noise-free simulation -- original excitable medium simulation with filaments

The movie video\_rounding1\_noise0\_k3.avi was made from the simulation when one decimal place was allowed, no noise was included, $g$ = 28, $maxstate$ = 200, $k_{1} = 3$ and $k_{2} = 3$. In this case, similarly as in the case when $k_{1} = 2$ and $k_{2} = 2$ it is sufficient to have one non-zero point to start ignition. This further illustrates the conclusion depicted in~Figure 4. In the movie, every 30th simulation step is shown with the frequency of 100 frames per second.

   \item Supplementary Material 5: Video of the simulation at optimal noise

The movie video\_rounding1\_noise\_appropriate.avi was made from the simulation at $k_{1} = 3$ and $k_{2} = 3$, 10 decimal places were allowed and  optimal levels of noise for each process, i.e., 0.00, 0.12, and 0.30 for process 1, 2a, and 2b, respectively. In the movie, every 100th simulation step is shown with the frequency of 100 frames per second.

   \item Supplementary Material 6: Video of the noise-free simulation at different $g/maxstate$ ratios

Files video\_rounding0\_noise0\_g1\_levels2000.avi, video\_rounding0\_noise0\_g280\_levels2000.avi, and video\_rounding0\_noise0\_g1000\_levels2000.avi show results of simulations at $k_{1} = 3$ and $k_{2} = 3$, $maxstate = 2000$, $noise = 0$, and $g$ as noted in the file name. Simulations exhibit distinct state space trajectories. In the movie, every 30th simulation step is shown with the frequency of 100 frames per second.

   \item Supplementary Material 7: Netlogo 5.2 model of the B--Z reaction as an excitable medium
\end{itemize}

\end{suppinfo}


\end{document}